\documentclass[prd,twocolumn,nofootinbib,superscriptaddress,showpacs]{revtex4}
\usepackage[dvips]{color}
\usepackage{url,epsfig,graphicx,amssymb,amsmath,axodraw}

\newcommand{\be}{\begin{equation}}
\newcommand{\ee}{\end{equation}}
\newcommand{\ba}{\begin{eqnarray}}
\newcommand{\ea}{\end{eqnarray}}

\newcommand{\E}{{\rm e}}

\def\cE{{\cal E}}
\def\cF{{\cal F}}
\def\L{{\cal L}}
\def\M{{\cal M}}
\def\D{{\cal D}}

\def\eps{\varepsilon}
\def\theta{\vartheta}
\def\sigv{\langle \sigma v\rangle}
\def\lsim{\raise0.3ex\hbox{$\;<$\kern-0.75em\raise-1.1ex\hbox{$\sim\;$}}}
\def\gsim{\raise0.3ex\hbox{$\;>$\kern-0.75em\raise-1.1ex\hbox{$\sim\;$}}}
\def\ap{\approx}

\def\Tr{{\rm Tr}}
\def\i{{\rm i}}
\def\d{{\rm d}}

\def \slash#1{\centeron{$#1$}{$/$}}
\def\centeron#1#2{{\setbox0=\hbox{#1}\setbox1=\hbox{#2}\ifdim
   \wd1>\wd0\kern.5\wd1\kern-.5\wd0\fi
   \copy0\kern-.5\wd0\kern-.5\wd1\copy1\ifdim\wd0>\wd1
   \kern.5\wd0\kern-.5\wd1\fi}}

\begin{document}

\title{On the role of electroweak bremsstrahlung for indirect dark matter signatures}

\author{M.~Kachelrie\ss}
 \affiliation{Institutt for fysikk, NTNU, N--7491 Trondheim,
  Norway}

\author{P.~D.~Serpico}
\affiliation{Physics Department, Theory Division,
CERN, CH--1211 Geneva 23, Switzerland}
\affiliation{LAPTH, UMR 5108, 9 chemin de Bellevue - BP 110, 74941 Annecy-Le-Vieux, France}

\author{M.~Aa.~Solberg}
 \affiliation{Institutt for fysikk, NTNU, N--7491 Trondheim,
  Norway}

\date{October 30, 2009}

\begin{abstract}
Interpretations of indirect searches for dark matter (DM) require theoretical 
predictions for the annihilation or decay rates of DM into stable particles 
of the standard 
model. These predictions include usually only final states accessible
as  lowest order tree-level processes, with  electromagnetic bremsstrahlung
and the loop-suppressed two gamma-ray line as exceptions. We show that this 
restriction may lead to severely biased results for DM tailored to produce 
only leptons in final states and with mass in the TeV range. 
For such models, unavoidable electroweak bremsstrahlung of $Z$ and $W$-bosons
has a significant influence both on the branching ratio and the spectral shape
of the final state particles. We work out the consequences 
for two situations: Firstly, the idealized case where DM  annihilates 
at tree level with 100\% branching ratio into neutrinos. For a given cross 
section, this leads eventually to ``minimal yields'' of photons, electrons, 
positrons and antiprotons. Secondly, the case where the only allowed 
two-body final states are 
electrons. The latter case is typical of models aimed at fitting cosmic ray 
$e^{-}$ and $e^{+}$ data. We find that the 
multimessenger signatures of such models can be significantly modified with 
respect to results presented in the literature.
\end{abstract}
\pacs{95.35.+d, 
95.85.Pw, 
98.70.Sa
}
\maketitle

\section{Introduction}
Despite the numerous cosmological and astrophysical indications for the 
presence of nonbaryonic dark matter (DM), the particle nature of DM has yet 
to be identified. One strategy towards DM ``detection" is to search
for its self-annihilation (or decay) products in our Galaxy, provided that 
the annihilation cross section $\sigv$ (or the decay rate) is large enough. 
Restricting the space of candidates to weakly interacting massive thermal
relics, $\sigv$ (at freeze-out) is fixed by the DM abundance while
the DM mass $m_X$  lies in this class of models typically within one or two 
orders of magnitude off the weak scale $m_W$. Nevertheless, considerable 
model-dependence remains because of our ignorance of the final states 
produced in the annihilation process, which vary from model to model.

Under the sole hypothesis that massive dark matter annihilates into 
standard model particles, a few years ago an interesting conservative upper
bound on $\sigv$ was derived using data on the least detectable final states, 
namely neutrinos~\cite{Beacom:2006tt,Yuksel:2007ac}. (Similar considerations 
apply to decaying particles~\cite{PalomaresRuiz:2007ry}, although we will not 
mention this further.) It is natural to ask if these conservative bounds can 
be improved or consolidated further. It was shown previously in 
Ref.~\cite{Berezinsky:2002hq} that electroweak bremsstrahlung  leads
to a break-down of perturbation theory and a non-negligible branching ratio 
in electromagnetic channels for $m_X\gg m_W$, even if at tree level 
DM couples only neutrinos. This argument was used then in 
Refs.~\cite{Kachelriess:2007aj,Bell:2008ey} to derive  constraints on
the DM annihilation cross sections from diffuse gamma-ray observations, that
turned out to be similarly restrictive as the original one from 
Ref.~\cite{Beacom:2006tt,Yuksel:2007ac}.

More recently, the PAMELA collaboration published data showing an ``anomalous 
growth'' of the cosmic ray positron fraction~\cite{Adriani:2008zr}, while
the measured antiproton fraction agrees with expectations from simples 
models~\cite{Adriani:2008zq}. Additionally, new data on the overall electron plus 
positron spectrum were presented, most notably from the Fermi space 
telescope~\cite{Abdo:2009zk,Grasso:2009ma}. These data have prompted a 
plethora of models trying to explain the data by engineering relatively heavy 
($m_X\agt 1\,$TeV) DM candidates with large annihilation cross sections 
(or decay rates) and small or vanishing branching ratios (br's) into hadronic final 
states. In the analysis of these models, the role of radiated $W,Z$ bosons and their 
phenomenological impact has been generally ignored.

The purpose of this article is to revisit the topic of  electroweak bremsstrahlung effects
with several goals in mind. First of all, albeit the qualitative conclusions of 
Refs.~\cite{Kachelriess:2007aj} and \cite{Bell:2008ey} agree, 
their results differ quantitatively. Here, we repeat this calculations
analytically and cross-check them against numerical results using
Madgraph. Second and more important for its recent phenomenological appeal, 
similar effects also arise when the tree-level final state is a charged lepton 
$\ell^{+} \ell^{-}$, and therefore we present an analogous calculation for an 
$e^{+} e^{-}$ pair as final state.
Third, besides photons, the fragmentation of the emitted $W$  and $Z$ bosons
produces also electrons and positrons, protons and antiprotons,  as well as neutrinos
and antineutrinos (hereafter, simply ``neutrinos''). Hence we can use observations of
different cosmic species to derive complementary constraints 
on DM annihilations. Finally, we include in the present analysis new data 
on the antiproton~\cite{Adriani:2008zq} and positron 
fraction~\cite{Adriani:2008zr} from the PAMELA satellite
as well as on diffuse gamma-rays from the Fermi Gamma-ray Space Telescope~\cite{MA}. 

This article is organized as follows: In Sec.~ \ref{analytical} we review
first some general considerations about the favoured DM annihilation modes
and then our analytical calculation of the branching ratio for electroweak
bremsstrahlung. In Sec.~\ref{numerical} we describe the spectra 
of secondaries found numerically, while Sec.~\ref{constraints} is devoted 
to the constraints from present observations. We conclude in 
Sec.~\ref{conclusions}.

\section{Analytical calculation of b.r.'s into $W,Z$.}\label{analytical}

\subsection{General considerations}
It is interesting to ask oneself under which circumstances it is possible to produce unsuppressed or
even dominant final states consisting of (possibly light or massless) leptons.
In the following, for the sake of self-consistency, we elaborate on some of the considerations reported  in the Appendix of~\cite{Bell:2008ey} and in~\cite{Cirelli:2008pk}. 

On general grounds, the 
$L$-th partial wave contribution to the annihilation rate of two heavy, 
non-relativistic particles moving with relative velocity $v$ is suppressed 
as $v^{2L}$. The virial velocity in our Galactic halo is only $v\sim10^{-3}$.
Typically, only the 
$L=0$ partial wave results in observable rates for indirect DM detection 
today. Admitting $L=1$ final states, terms proportional to $v^2$ and 
terms of similar magnitude that are chirally suppressed as 
$\propto (m_f/m_X)^2$, where $m_f$  denote
the mass of fermions, enter the 
corresponding annihilation rate: This is notably the case for annihilations 
via an axial vector current,  $J^{PC}=1^{++}$. (We use here the spin-parity 
notation $J^{PC}$, where $J$ is the total---orbital plus internal---spin 
quantum number, $P$ the parity and $C$ the charge-parity).

Such a scenario has two classes of phenomenological problems: First, any 
observable effect now should produce huge effects at early times, when 
$v$ was larger. Second, multi-body final states, e.g.\ with e.m. or weak 
boson radiation emitted from one fermion, proceeds unsuppressed as shown 
for the case of the axial vector mode in Ref.~\cite{VSRold}.
In this case, bounds from photon and cosmic ray antimatter searches typically 
rule-out these models. This provides a general argument to focus on 
$L=0$-modes only, as we do in the following.

For a spin-1/2 Majorana fermion candidate, the parity is $P=(-1)^{L+1}$ and 
the charge parity state $C=(-1)_{\rm orbital}^{L}(-1)_{\rm spin}^{S+1}(-1)_{\rm anticomm.}=(-1)^{L+S}$, where the spin $S=0,1$ for a fermion pair. 
Since $L=0$ is the only acceptable choice, it follows that $P=-1$. Also, 
the Majorana nature of the particles implies that it must be even under $C$, 
hence $S=0$. The only state selected is thus the pseudoscalar $0^{-+}$.

Another class of popular DM candidates are scalars. Again, since we are in 
the non-relativistic limit, the only unsuppressed annihilation state is the 
scalar one, with $L=0\to P=-1$.  Either a fundamental DM scalar or 
pseudoscalar would lead to a scalar singlet. The state is clearly even under 
$C$-symmetry. Thus,   $J^{PC}=0^{++}$, i.e.\ only a scalar is allowed.
A spin-0 two body particle states is also allowed in case of Dirac and vector 
particles as DM candidates, although it is then not the unique choice leading
to $L=0$ annihilations~\cite{Cirelli:2008pk}.

In summary, for the following calculations we can adopt an effective field 
theory approach and define the initial state as 
$\D$ coupling with SM neutrinos $\nu$  as
$-\i\lambda\,\D \bar{\nu}\nu$  or the analogous pseudoscalar coupling
$-\i\lambda\,\D \bar{\nu}\gamma^5\nu$.  
Basically without loss of generality, this is a viable way to obtain 
on a phenomenological level a tree-level coupling to neutrinos only. 
Note that the exact flavour-structure of the coupling is irrelevant, 
since the two large neutrino mixing angles lead after oscillations 
to a $1:1:1$ mixture of flavours. 
A similar coupling will be adopted for the electron final state case
as well.

\subsection{Explicit calculation}

\begin{figure*}[!ht]
\includegraphics[angle=0,width=0.6\textwidth]{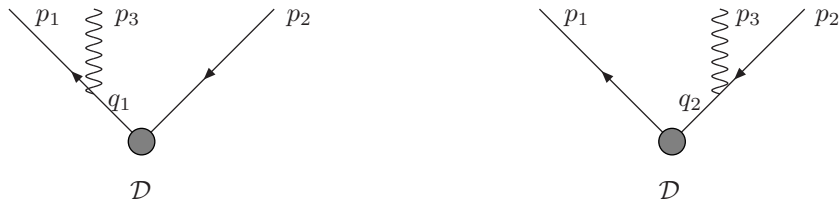}
\caption{Feynman diagrams for the process $\D\to \bar\nu\nu Z$,
left $q_1=p-p_2$ and right $q_2=p-p_1$, with $p$ being the $\D$-state four-momentum.}\label{Feyn}
\end{figure*}%

As discussed above, we can replace e.g.\ the annihilation process $\bar X X\to \bar\nu\nu Z$ 
with the decay 
$\D\to \bar\nu\nu Z$ choosing $m_\D=2m_X$. In particular, this
replacement becomes exact, if the annihilation process is mediated by
a scalar particle,
\be
 R_Z = \frac{\sigma(\bar XX\to \bar\nu\nu Z)}{\sigma(\bar XX\to \bar\nu\nu)}
     = \frac{\Gamma(\D\to \bar\nu\nu Z)}{\Gamma(\D\to \bar\nu\nu)} \,.
\ee
To be specific, we choose a scalar coupling $\L_I=-\lambda\D\bar\nu\nu$
between neutrinos and $\D$. The tree-level decay rate of $\D$ in one massless
neutrino flavour is then $\Gamma(\D\to\nu \bar{\nu})=\lambda^2 m_\D/(8\pi)$.
Note that this is a factor two smaller than the decay rate used in 
Ref.~\cite{Bell:2008ey}. 
Also note that in the above expression $\nu$ is implicitly assumed to be a Dirac field, hence 
equal populations of active left-handed neutrinos $\nu_L$ and ``sterile'' right-handed
states $\nu_R$ are produced. 
While there exists no evidence for the existence of $\nu_R$'s up to now, 
this choice leads to the most conservative bounds: Since the ``sterile'' 
right-handed neutrino states $\nu_R$ do not participate in electroweak 
interactions, the br's of the radiative channels would be enhanced 
restricting the tree-level coupling to $\nu_L$ only. 
  
The two diagrams describing the electroweak Bremsstrahlung process 
 and our notation for the momenta are shown in 
Fig.~\ref{Feyn}. The corresponding Feynman amplitudes are 
\begin{align}
 \M_1 = \bar u(p_1)\:\frac{-\i g}{2c_W}\gamma^\mu P_L\: \frac{\i}{\slash q_1}\: \eps_\mu (p_3) \:(-\i\lambda)\: v(p_2)\,, 
\\
 \M_2= \bar u(p_1) \: (-\i\lambda)\:\frac{\i}{\slash q_2}\:\frac{-\i g}{2c_W}\gamma^\nu P_L \: \eps_\nu (p_3)\: v(p_2)\,,
\end{align}
where we have neglected neutrino masses, $m_\nu=0$, as always in the 
following. Furthermore, $g$ is the weak coupling constant, $c_W=\cos\theta_W$
and $P_{L/R}=(1\pm\gamma^5)/2$ are the left/right-chiral projection operators.
Using as short-hand notation for the polarization tensor of
a massive gauge boson $P_{\mu\nu}=-g_{\mu \nu}+p_{3\mu}p_{3\nu}/m_Z^2$ 
as well as $K_i=[(g\lambda)/(2c_Wq_i^2)]^2$, we obtain summing over spins
\be
 \overline{\M_1\M_1^\ast} = K_1 P_{\mu \nu} 
 \Tr(\slash p_1 \gamma^\mu P_L \slash q_1 \slash p_2 \slash q_1 \gamma^\nu)
 = K_1 P_{\mu \nu} B^{\mu \nu}_1 \,.
\ee
The interference term vanishes for $m_\nu=0$ and thus we can set
$|\M|^2=2|\M_1|^2$ in the calculation of the decay rate. Evaluating
the trace in the rest-frame of $\D$ gives  with $p=(m_\D,0)$ 
\ba
\lefteqn{\frac{P_{\mu\nu} B^{\mu\nu}_1}{q_1^4} = 
 \frac{2 \left(m_\D^2+m_Z^2-m_{12}^2\right)}{m_\D^2+m_Z^2-m_{12}^2-m_{23}^2}}
\\
& -& \frac{2m_\D^2 m_Z^2}{\left(m_\D^2+m_Z^2-m_{12}^2-m_{23}^2\right)^2}
 +\frac{m_{12}^2}{m_Z^2}-2 \,,
\nonumber
\ea
where we introduced $m_{ij}^2=(p_i+p_j)^2=(p-p_k)^2$~\cite{West}.

The differential decay rate for a general $1\to 3$ decay process,
\be
 \d\Gamma = 
 \frac{1}{(2\pi)^3}\frac{1}{32m_\D^3} |\overline{\M}|^2 \d m_{12}^2  \d m_{23}^2 \,,
\ee
can be integrated exactly using the limits~\cite{PDG}
\be
 (m_{23}^2)_\text{min}^\text{max} = \frac{1}{2} \left[
    m_\D^2 \pm \left( \Delta^2 - 2\Sigma m_\D^2+m_\D^4 \right)^{1/2}
     + \Delta \right]
\ee
with
\be
 \Delta = m_Z^2 - m_{12}^2 \qquad{\rm and}\qquad
 \Sigma = m_Z^2 + m_{12}^2 \, .
\ee
The resulting ratio $R_Z$ is
\be
 R_Z =
 \frac{g^2}{384\pi^2 c_W^2} y_Z \Big[  1-\frac{9}{y_Z}-\frac{9}{y_Z^2}
       +\frac{17}{y_Z^3}  + 
\left( \frac{18}{y_Z^2}+\frac{6}{y_Z^3}\right) \ln y_Z \Big] \,,
\ee
with $y_Z\equiv p^2/m_Z^2=4\,m_X^2/m_Z^2$. The ratio $R_W$ follows from $R_W(y_W)=2c_W^2 R_Z(y_W)$. Our ratios have the
same dependence on $y_{Z,W}$ as those found by the authors of 
Ref.~\cite{Bell:2008ey}, but are a factor four larger. A factor of two is 
explained by the difference in the tree-level decay width, while the other, given the agreement
of our analytical results with numerical ones presented in the following, is attributed to a missing algebraic 
factor in the calculation of~\cite{Bell:2008ey}.  

For the electron final state case, the ratio $R_W$ writes similarly as above, but the different
structure of the coupling with $Z$ (involving also $P_R$) does not lead to a similar simple
expression for $R_Z$; in particular, interference terms do not vanish in the unitary gauge. The latter
contribution has been thus calculated only numerically.

In our calculation of $R_Z$ we have included only final state radiation (FSR) 
neglecting virtual state radiation (VSR) ``from the decay vertex'' as well
as initial state radiation (ISR).
The latter two can be only calculated within specific models or assuming a 
separation of scales such that an effective theory approach can be used.
The separation between three classes of bremsstrahlung is gauge-dependent 
and thus strictly 
speaking meaningless. However, in cases where one of the three radiation 
mechanisms dominates and no cancellations are present, this separation is 
useful. For the electromagnetic case, the relative importance of VSR and FSR
is discussed e.g.\  in Ref.~\cite{VSRold,Bringmann:2007nk,Barger:2009xe}. 
In some cases, VSR is important 
since, depending on the exact spin structure of the interaction and the
particles involved, the helicity suppression present for certain two body 
final states can disappear. This is the opposite limit to the one we are 
discussing here, since we are working within
the {\em ansatz\/} of dominant neutrino or electron two-body final states. 
For our purpose, it is reasonable to conclude that neglecting the 
(model-dependent) ISR and VSR provides at most a slightly conservative 
evaluation of the visible final state channels.

Also, it has been shown in~\cite{Dent:2008qy} that, in addition to the $Z$ and 
$W$-strahlung processes, loop-induced branching ratios into charged leptons (and, we note, more generally into charged fermions including quarks) are unavoidable in these scenarios. Similarly to the VSR, these effects depend on the UV completion of the model and we do not discuss this further. Typically, one expects these branching ratios to be at  the percent level, and thus mainly important for relatively
light DM particles, $m_X\lsim m_W$. In any case, loop processes as additional
source of photons and anti-matter would strengthen the limits derived in the 
following.

Let us discuss now the behavior of the bremsstrahlung amplitudes in the limit 
$m_\D\gg m_Z$. 
The similar calculation of the three-jet rate in QCD with the emission of a 
massive gluon leads to $R_{\rm QCD}\propto \ln^2(m_\D^2/m_g^2)$, and in
this process the longitudinal gluon does not contribute as it is coupled to 
a conserved current. By contrast, the axial vector current is not conserved
and it is indeed the longitudinal polarization of the electroweak gauge bosons 
that is responsible for the quadratic mass dependence, 
$R_i\propto m_\D^2$~\cite{Ciafaloni:2009tf}.
This difference is also the main reason for the discrepancy between the results
of Ref.~\cite{Kachelriess:2007aj} and~\cite{Bell:2008ey}. In the former work, 
the probability  for electroweak bremsstrahlung was calculated  using 
the Feynman-'t Hooft gauge and as tree level process 
$\bar\nu'\nu'\to\bar\nu\nu$. Hence, additionally to the differences expected
from using a different current and coupling structure mediating the 
annihilations, the slower rise of the br's (which also holds in SUSY, see
below) found in~\cite{Kachelriess:2007aj} can be attributed to the absence 
of the longitudinal polarization of the gauge bosons.

Unitarity requires e.g.\ that the annihilation cross section behaves as
$\sigv\propto m_X^{-2}$. While this  bound is respected by the tree-level
cross section, the $R_i^2\propto m_\D^2$ dependence leads to a violation
of perturbative unitarity if an additional $W$ or $Z$ is emitted. 
The decoupling property of supersymmetry~\cite{dec}
guaranties that such a quadratic term is absent, even in the presence
of soft masses $\gg m_W$. An explicit calculation in
the MSSM show that the ratio of e.g.\ 
$\sigma(\chi\chi\to e^+e^-Z)/\sigma(\chi\chi\to e^+e^-)\sim 0.03$ 
for a 10\,TeV neutralino. Hence the MSSM is an example for a scheme where
initial, virtual and final state radiation from all sub-processes in the 
$s$, $t$ and $u$ channel is arranged in such a way that their leading 
contributions cancel. 
On the other hand, in such a theory quarks and gauge bosons are produced at 
tree-level in two-body  final states, so that the yields
of secondaries other than leptons are typically large.

Similarly, the above considerations can be circumvented if the DM does not decay directly into
SM particles, rather trough some {\it light} state $Y$, so that $W,Z$ in the final state are kinematically inaccessible. These models however introduce other {\it light}, metastable degrees
of freedom, which violate from the beginning our ``effective theory'' approach. On the other hand, for a
chain $X\bar{X}\to n\,Y\to SM$, provided that $m_Y\geq m_Z$, one might still apply the above
arguments with the new ratios written now as $R_{Z,W}'=n\,R(m_Y^2/m_{Z,W}^2)$, although the distributions of secondaries will be different, of course.

\section{Numerical Results}\label{numerical}

\begin{figure}
\includegraphics[angle=0,width=0.5\textwidth]{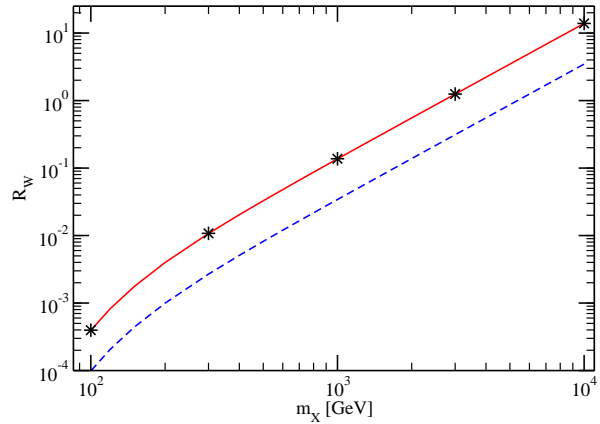}
\caption{Our analytical result for the ratio $R_W$ (solid line), compared to 
$R_W$ from ref.~\cite{Bell:2008ey} (dashed line) and the numerical results from 
Madgraph (points). \label{fig:Rw_comp}}
\end{figure}

We have calculated the total cross section $\bar XX\to \nu_e e^\pm W^\mp$
also numerically, using Madgraph~\cite{Alwall:2007st} and
allowing electroweak bremsstrahlung only as FSR. 
More precisely, we used the MSSM model, choosing $X$ as the lightest
neutralino and the scalar higgs $h$ as intermediate particle as well
as switching off all diagrams except the one corresponding to FSR.
We denote the ratio of the bremsstrahlung and the tree-level processes
again as $R_W$,
\be
  R_W =\frac{\sigma_3}{\sigma_2}=
 \frac{\sigma(\bar XX\to\nu_e e^+ W^-)+\sigma(\bar XX\to \bar\nu_e e^- W^+)}{\sigma(\bar XX\to \bar\nu_e\nu_e)} \,.
\ee
In Fig.~\ref{fig:Rw_comp}, we compare our analytical result for $R_W$
with the numerical results obtained with Madgraph, finding excellent
agreement.

Note that the perturbative results for the bremsstrahlung cross sections 
become unreliable already at $m_{X}\gsim 1$\,TeV. Above this energy, 
processes with $n>3$ (treating $W,Z$ as stable) particles in the final state 
become important and an electroweak cascade develops~\cite{Berezinsky:2002hq}. 
On the other hand, each individual sub-process is suppressed by a Sudakov 
factor compared to the perturbative result, avoiding a ``blowing-up'' of the 
total annihilation cross section. 

\begin{figure*}[!th]
\begin{center}
\begin{tabular}{cc}
\includegraphics[angle=0,width=0.5\textwidth]{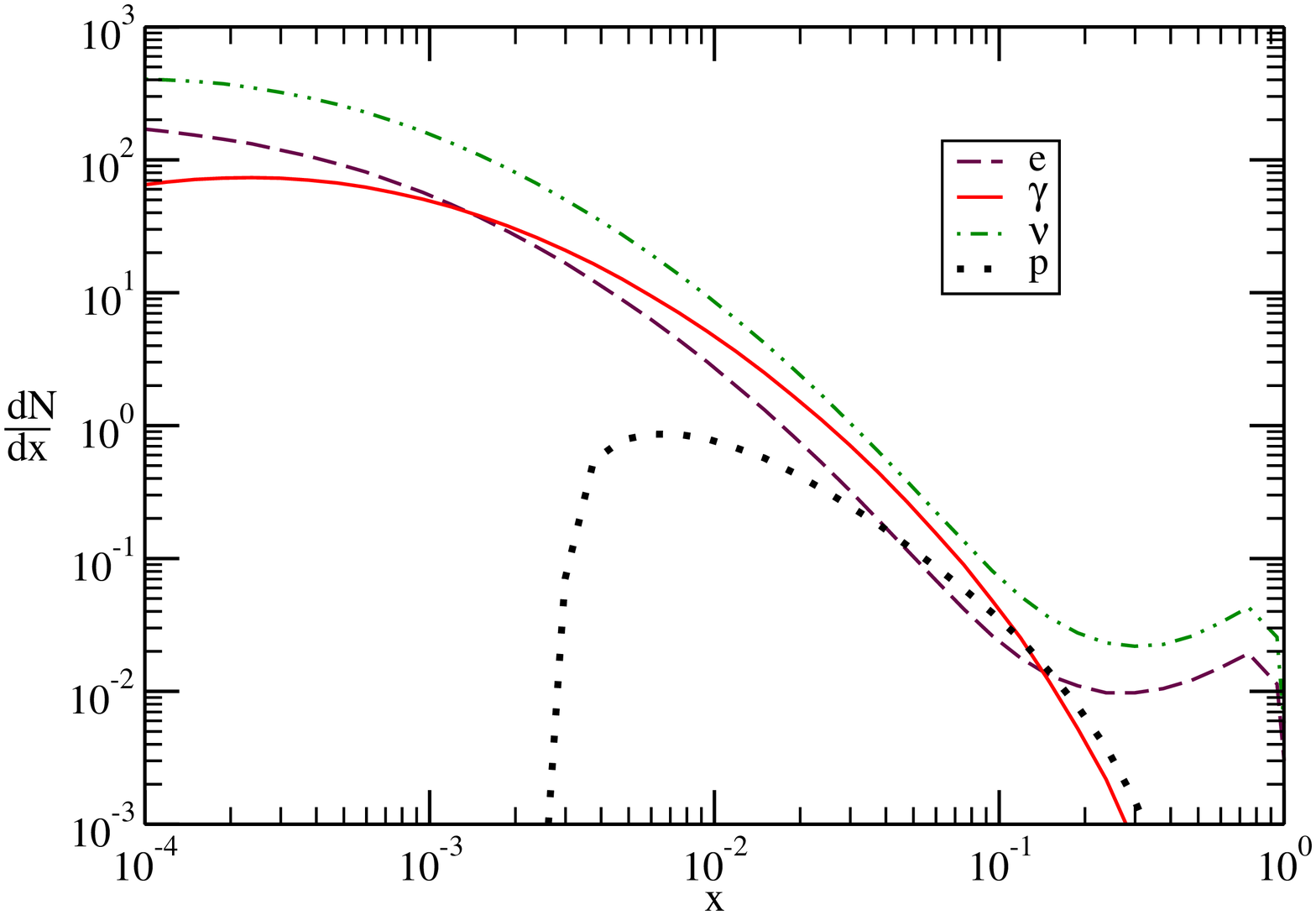}
&
\includegraphics[angle=0,width=0.5\textwidth]{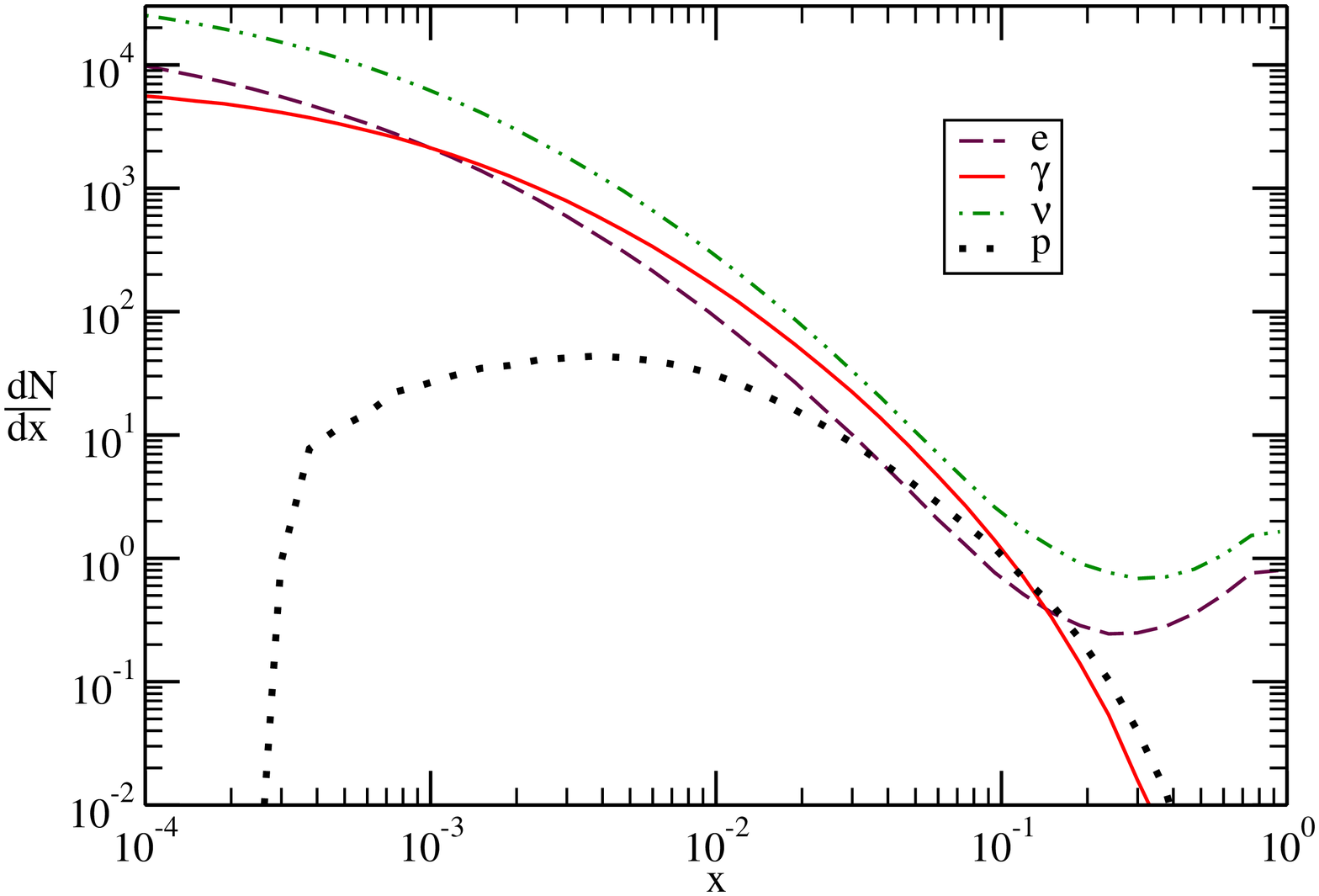}
\end{tabular}
\end{center}
\caption{The fragmentation spectra $\d N/\d x$ of electrons (long-dashed maroon), 
photons (solid red), neutrinos (dot-dot-dashed green), and protons (dotted black) 
in ``only neutrinos'' tree level annihilations for $m_X=300$\,GeV (left) and 
$m_X=3$\,TeV (right). The monochromatic neutrino ``spike'' at $x=1$ is not shown. 
Fermion labels indicate the sum of matter and antimatter.
\label{fig:nunu}}
\end{figure*}

In the next step, we feed the events generated by Madgraph into  
HERWIG++~\cite{HERWIG++} and generate the energy spectra of 
$e^\pm$, $\nu_i,\bar\nu_i$, $p$, and $\bar p$ produced as secondaries 
in $W$ and $Z$ decays. 
The obtained energy spectrum $\d N/\d x$ are normalized to the total cross 
section, $\sigma_{\rm tot}=\sigma_2+\sigma_3$,
\be
 \frac{\d N_i}{\d x} = \frac{f_i}{\sigma_{\rm tot}}\:
                   \frac{\d\sigma(XX\to i+{\rm all})}{\d x}
\ee
where $f_i$ is the multiplicity to produce particles of type $i$ with
energy $E=xm_X$. Therefore the evolution of $\d N/\d x$ with energy 
becomes much slower for $m_X\sim$ few TeV, when 
$\sigma_{\rm tot}\ap\sigma_3\gg \sigma_2$.
Since the shape of the fragmentation function $\d N_i/\d x$ changes only
logarithmically, the omission of $2\to n>3$ processes has only a minor
impact on  $\d N_i/\d x$. Moreover, the chosen normalization gives the
correct number of secondaries $N_i$ per annihilation also for 
$\sigma_3\gg\sigma_2$.
The resulting fragmentation functions $\d N/\d x$ as function of $x$ are 
shown for $m_X=300$\,GeV  and $m_X=3$\,TeV in Fig.~\ref{fig:nunu}. Note  
that i) there is an additional contribution 
$\delta(1-x)\,\sigma_2/\sigma_{\rm tot}$ which is not reported for
clarity and ii) only half of the neutrinos emitted at tree-level are 
``active'' ones, $\nu_L$.

In the same way as described above  for tree-level annihilations
into neutrinos, we have calculated also the total cross sections of 
$\bar XX\to e^+e^-$ and $\bar XX\to e^+e^-Z$, $\bar XX\to \nu_e e^\pm W^\mp$.
Additionally, we added  analytically photons from external electromagnetic 
bremsstrahlung,
\be
 \frac{\d N_\gamma}{\d x} = \frac{\alpha}{\pi} \, P_{\rm ff}(x) \,\ln\frac{s(1-x)}{m
_e^2} 
\ee
with the usual fermion-fermion-boson splitting function 
\be
 P_{\rm ff}(x)=\frac{1+(1-x)^2}{x} \,.
\ee

\begin{figure*}[!th]
\begin{center}
\begin{tabular}{cc}
\includegraphics[angle=0,width=0.5\textwidth]{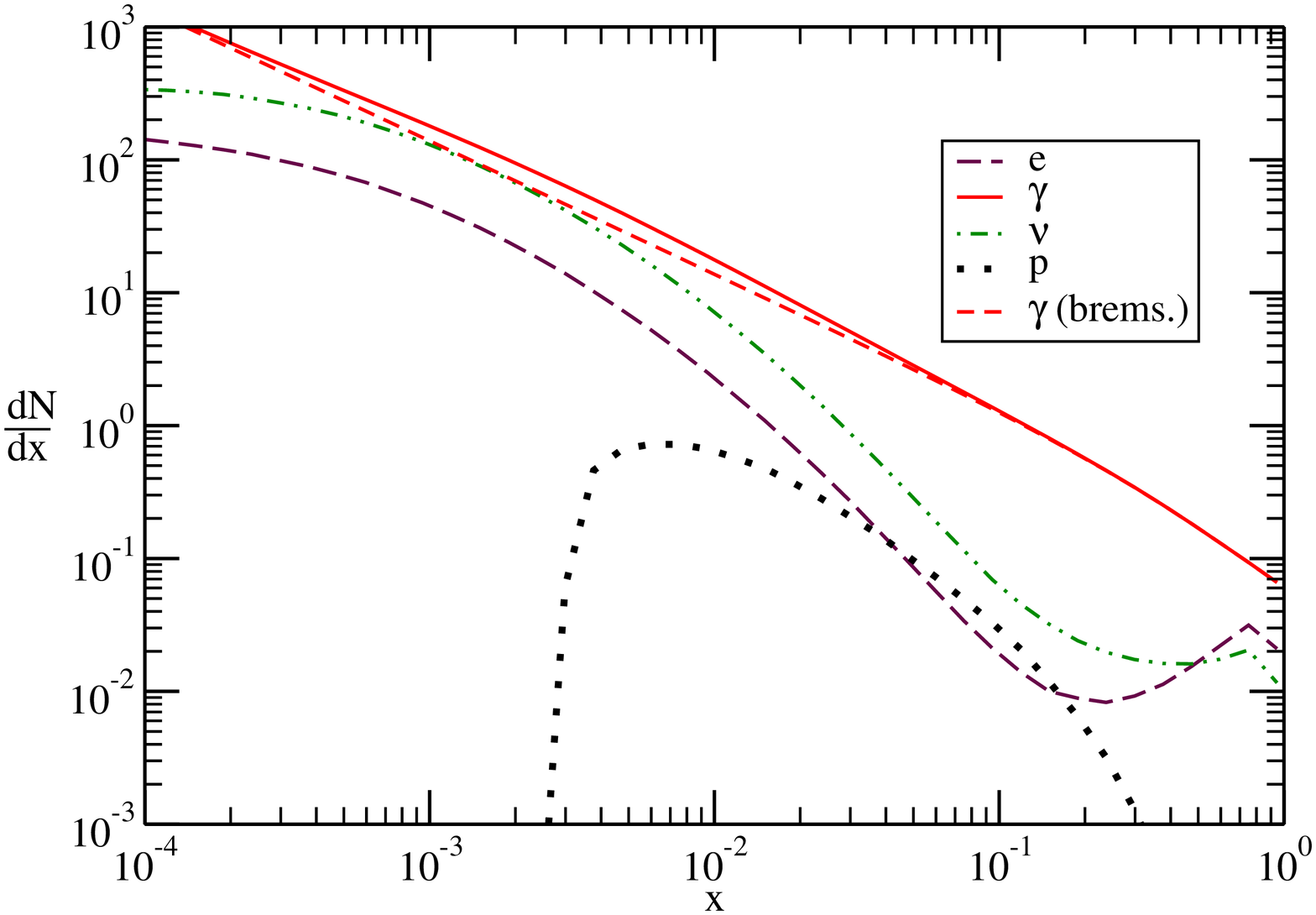}
&
\includegraphics[angle=0,width=0.5\textwidth]{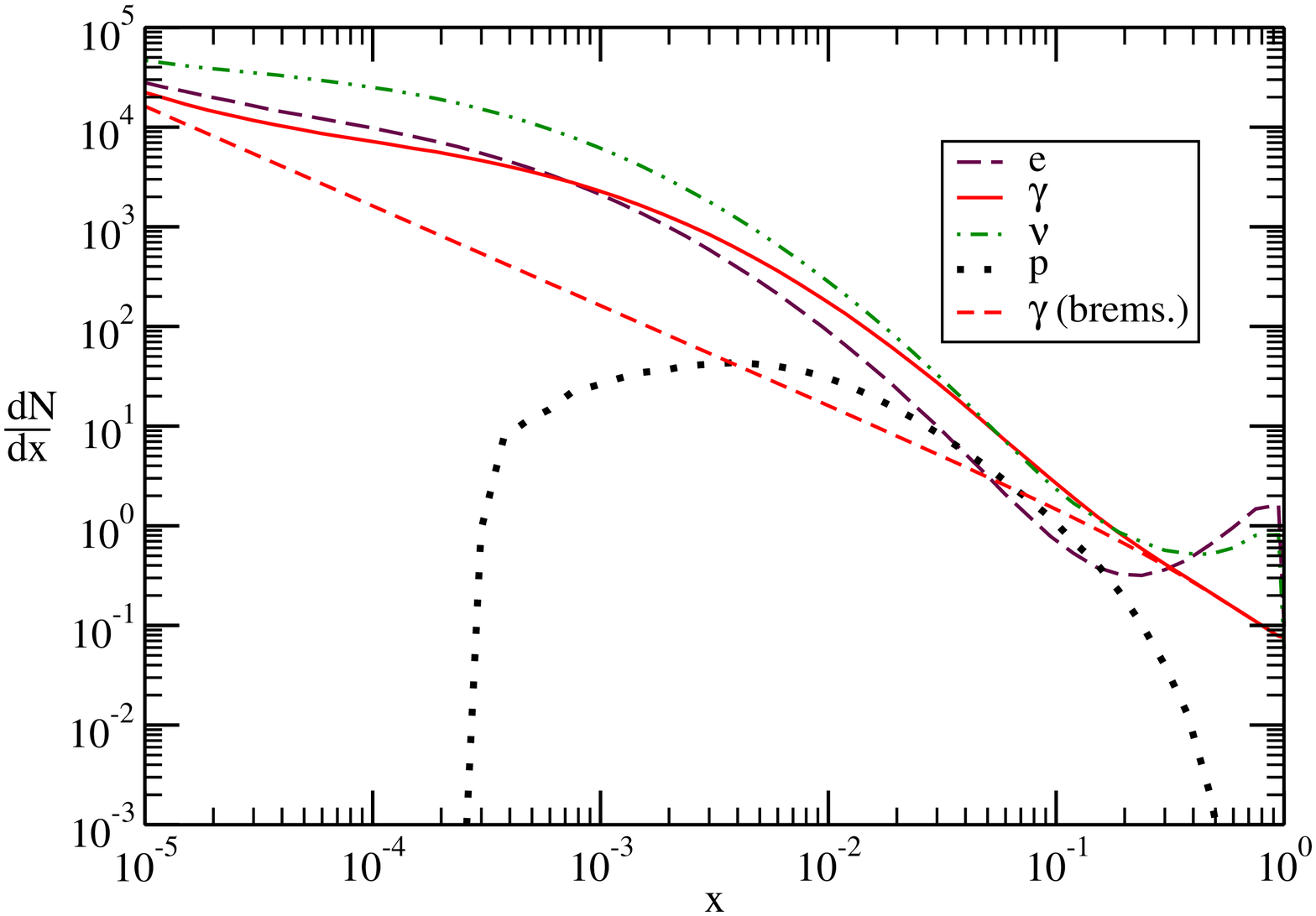}
\end{tabular}
\end{center}
\caption{The fragmentation spectra $\d N/\d x$ of electrons (long-dashed maroon), photons (solid red),
neutrinos (dot-dot-dashed green), and protons (dotted black) in ``only electron'' annihilations for $m_X=300$\,GeV (left) and $m_X=3$\,TeV (right). Also shown is the gamma-ray yield
from bremsstrahlung only (short-dashed red). The monochromatic
electron ``spike'' at $x=1$ is not shown. Fermion labels indicate the sum of matter and antimatter.
\label{fig:ee}}
\end{figure*}

The resulting fragmentation functions $\d N_i/\d x$  are shown for 
$m_X=300$\,GeV  and $m_X=3$\,TeV in Fig.~\ref{fig:ee}.
Note that the relative importance of protons is largest for $x\sim 0.1$ 
where it is comparable with the secondary electron flux.
Also, bremsstrahlung photons (absent for the neutrino case) provide the dominant
yield only for sub-TeV DM masses, while for heavy DM particles,
apart for the region $x\agt 0.1$ and the very small $x$, the photons from $W,Z$
fragmentation dominate.  As above, there is an additional contribution to the
$e^{-}+e^{+}$ case of $2\,\delta(1-x)\,\sigma_2/\sigma_{\rm tot}$ which is not reported for
clarity, but of course included in the constraints derived below. Note also 
that, in contrast to the neutrino case, both helicity states emitted
at tree-level contribute to the observable electron spectrum.

\section{Constraints}\label{constraints}
In this section we provide a first application of the calculations reported 
above, deriving bounds on $\sigv$ vs. $m_{X}$ from measurements of the diffuse gamma ray flux, 
the antiproton fraction, the electron plus positron flux, the positron fraction and the limits on Galactic
neutrino fluxes. We shall 
adopt a simple prescription, namely we consider a model as excluded when 
the DM contribution to the signal {\em alone\/} exceeds the maximal flux 
allowed by the data at one sigma. This is likely over-conservative
since it is known that other more mundane astrophysical sources contribute 
to or even dominate the fluxes.  For easing the comparison with other papers,  
we adopt for the smooth DM mass density $\rho_{\rm sm}$ a fiducial 
Navarro-Frenk-White profile~\cite{NFW}
\begin{equation} 
 \rho_{\rm sm}(r)=\rho_\odot\left(\frac{r_\odot}{r}\right)
 \left(\frac{r_\odot+a}{r+a}\right)^2,
\end{equation}
where $a= 20\,$kpc is the characteristic scale below which the profile scales as $r^{-1}$. Following the new determination in~\cite{Catena:2009mf},  we choose $\rho_\odot =0.39\,$GeV/cm$^3$ as the DM density at the solar distance from the  (Galactic center) GC; 
the latter is estimated to be $r_\odot\approx 8.33\,$kpc~\cite{Gillessen:2008qv}. Since we shall refer to high-latitude fluxes (for the photon case) or diffuse signals
(for the charged particles cases) the exact DM 
profile towards the GC is not essential. For neutrinos, we
shall limit this dependence adopting a quite large cone size (see below).

Note that additional bounds can be derived by focusing on more specific 
location (as the GC) and/or by looking at other channels produced as 
secondaries of leptonic energy losses (inverse Compton,
synchrotron radiation,\ldots). It is beyond the scope of this paper to 
provide an exhaustive account of indirect bounds. Here we only note that 
most of them depend more on the properties of the Galactic medium, and that 
further constraints can only strengthen  the  results presented here.
 
\subsection{Gamma-rays}
The differential flux of photons from (self-conjugated) dark matter annihilations is
\begin{equation}
\Phi_{\gamma}(E,\psi)=\frac{\d N_\gamma}{\d E}\,\frac{\sigv\,\rho_\odot^2}{8\,\pi\,m_X^2}\,\int_{\rm
l.o.s.} \d s\,\left(\frac{\rho_{\rm sm}[r(s,\psi)]}{\rho_\odot}\right)^2\,, \label{Ism}
\end{equation}
where $r(s,\psi)=(r_\odot^2+s^2-2\,r_\odot\,s\cos\psi)^{1/2}$,
$\psi$ is the angle between the direction in the sky and the
GC  and $s$ the distance from the Sun along the
line-of-sight (l.o.s.). In terms of the galactic latitude $b$ and longitude $l$, one has
$
\cos\psi=\cos b\cos l\,.
$
Particle physics enters via the DM mass
$m_X$, the annihilation cross section $\sigv$, and
the photon differential energy spectrum $\d N_\gamma/\d E$
per annihilation.
Since here we are focusing on the galactic diffuse
emission rather than that from the GC, the residual uncertainties
which are introduced through the choice of the DM profile are negligible for our discussion. We shall compare the calculated
flux with the high-latitude residual ``isotropic'' emission whose preliminary
data have been presented by the Fermi-LAT collaboration~\cite{MA}, whose upper limit at $\sim$ 1 $\sigma$
in the range 0.1--50 GeV can be roughly represented by $E^2 \d N/\d E\simeq  1.5\times 10^{-2}(E/0.1\,{\rm GeV})^{-0.45}\,{\rm GeV}\, {\rm m}^{-2}{\rm  s}^{-1}\,{\rm sr}^{-1}$.

\subsection{Antiprotons}
Accounting for diffusion, the flux of antiprotons at Earth is isotropic  to a high accuracy. By neglecting energy losses and the so-called ``tertiary'' component (which is mostly relevant at low-energies) the flux can be written similarly to Eq.~(\ref{Ism}) as 
\begin{equation}
\Phi_{\bar{p}}(E)=\frac{\d N_{\bar{p}}}{\d E}\,\frac{\sigv\,\rho_\odot^2}{8\,\pi\,m_X^2}\,\cF_{\bar{p}}(E)\,, \label{Ipbar1}
\end{equation}
where $\d N_{\bar{p}}/\d E$ is now the antiproton injection spectrum per annihilation, while
the integral along the line of sight is effectively replaced by a function $\cF_{\bar p}(E)$ which encodes
the dependence of the flux from astrophysical parameters (see~\cite{Donato:2003xg} for a derivation of the above formula and an explicit expression of $\cF_{\bar {p}}$ ). 
For the present illustrative purposes, we adopt
the fit of this function calculated numerically in~\cite{Cirelli:2008id}, for the reference NFW model and the ``intermediate'' choice of propagation parameters (see~\cite{Cirelli:2008id} for details).

In order to compare with the $\bar{p}/p$ ratio provided in~\cite{Adriani:2008zq}, we normalize the
above calculated flux to the observed proton flux which we take from~\cite{pdg}
to be
\begin{equation}
\Phi_{p}^{\rm obs}(E)= 1.4\times 10^4 (E/{\rm GeV})^{-2.7}\,\,{\rm GeV}^{-1}\, {\rm m}^{-2}{\rm  s}^{-1}\,{\rm sr}^{-1}\,, \label{Ipbar2}
\end{equation}
where we accounted for a proton contribution of 79\% in the overall cosmic ray 
flux at the energies of interest.

\subsection{Electrons and Positrons}
Compared to the above case, the main difference concerning the calculation of the  flux of $e\equiv (e^{-}+e^{+}$) at the Earth starting from the injected parameters is due to the relevance of energy losses. A semi-analytical form can be derived~\cite{Hisano:2005ec,Delahaye:2007fr}, which yields
\begin{equation}
\Phi_{\E}(E)=\frac{\d N_{\E}}{\d E}\,\frac{\sigv\,\rho_\odot^2}{8\,\pi\,m_X^2}\cF_{\E}(E)
\,, \label{Ipos}
\end{equation}
with
\begin{equation}
\cF_{\E}(E)
=\frac{1}{\frac{\d N_{\E}}{\d E}(E)\,b(E)}\int_E^{m_X} \d\cE\frac{\d N_{\E}}{\d E}(\cE)\,h_{\E}(\cE)\,.
\label{lpos2}\end{equation}
In the above equation, $b(E)\approx 10^{-16}\,(E/{\rm GeV})^2\,{\rm GeV\,s^{-1}}$ encodes energy losses and  the function $h_{\E}(E)$ depends on halo parameters as well as propagation ones. As for the case of antiprotons,
we adopt the fit for this function provided in~\cite{Cirelli:2008id}  for the NFW case and intermediate propagation parameters. Note that the positron flux amounts to half of the flux reported above.

From the overall electron plus positron spectrum we know from the Fermi measurement that~\cite{Abdo:2009zk,Grasso:2009ma}
\ba  \label{fermi}
\lefteqn{\Phi_{\E}^{\rm obs}(E)\equiv\Phi_{\E^{+}+\E^{-}}^{\rm obs}(E) =}
\\ &&
(175.40\pm 6.09)
\left(\frac{E}{{\rm GeV}}\right)^{-3.045\pm 0.008}\,
\frac{1}{{\rm GeV}\, {\rm m}^{2}\,{\rm  s}\,{\rm sr}}
\nonumber 
\ea
represents a good fit of the data between $\sim 20\,$GeV and 1 TeV. We shall then require that:
i) $\Phi_{\E}/(2\,\Phi_\E^{\rm obs})\leq f_{\E^{+}}^{\rm up}(E)$, where  $ f_{\E^{+}}^{\rm up}(E)$ is the upper value of the data-points presented by PAMELA at energies between $\sim 20$ and $\sim 100$ GeV. ii) That $\Phi_{\E}\lesssim \Phi_\E^{\rm obs,up}$, where  $\Phi_\E^{\rm obs,up}$ is the flux obtained for the upper value of the flux normalization  and the hardest spectrum reported in~Eq.~(\ref{fermi}).

\subsection{Neutrinos}
For neutrinos, we use the Super-Kamiokande bound~\cite{Desai:2004pq} requiring that the
induced flux of muon tracks from a cone of half-width 30$^\circ$ around the GC
is below 1.6$\times 10^{-14}$/s (we specifically use the slightly conservative prescription to use
only events induced by neutrinos above 10 GeV energy). 
Also, we assume a $1:1:1$ mixture of flavours, which holds within factors of order unity independently
of the production flavour ratios due to mixing. Finally, note that the width of the cone around the Galactic
Center is such that the dependence of the flux on the exact DM profile is marginal for our purposes (within a factor 2 at most, see~\cite{Yuksel:2007ac}).

\subsection{Results}
\begin{figure}
\includegraphics[angle=0,width=0.5\textwidth]{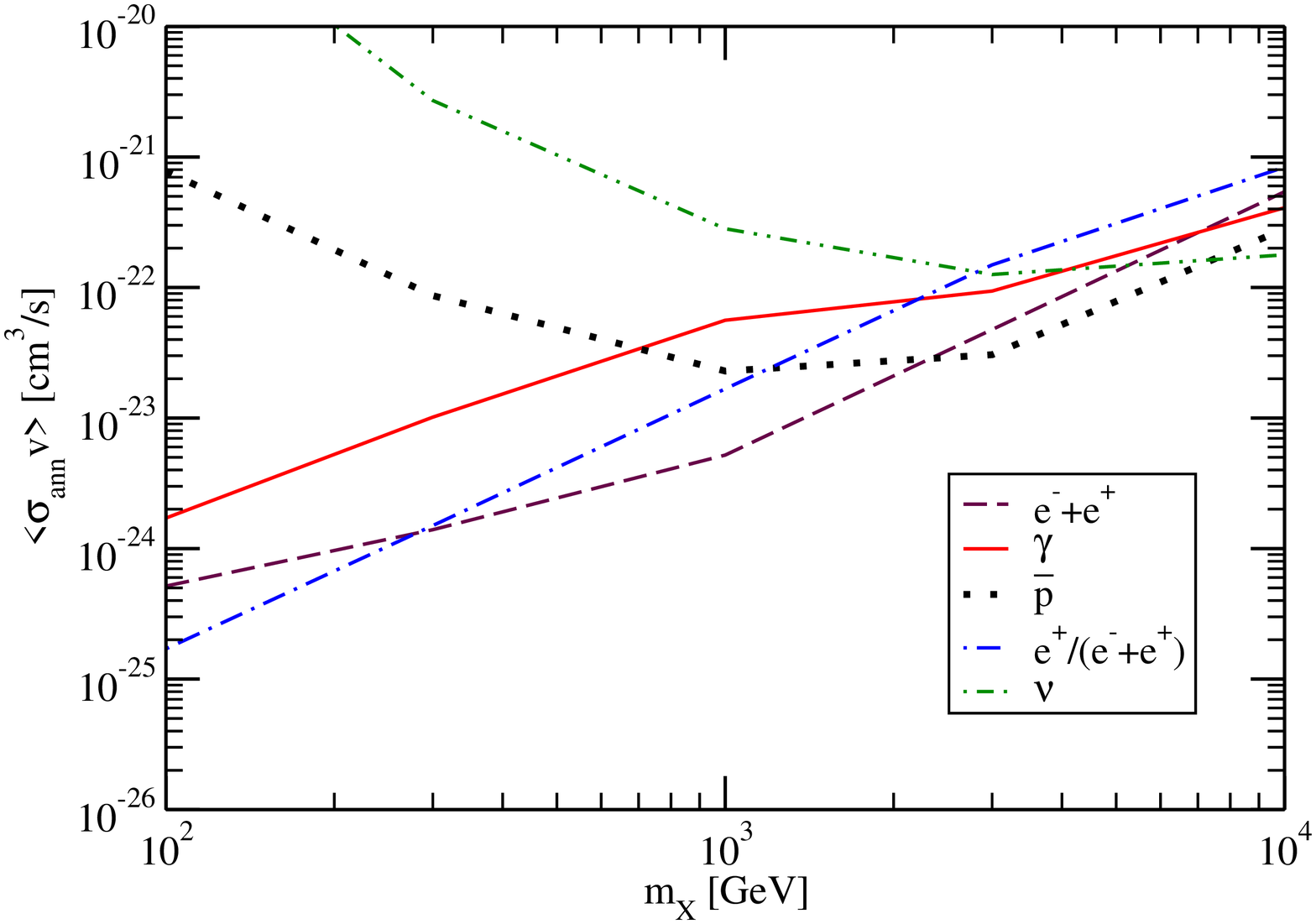}
\includegraphics[angle=0,width=0.5\textwidth]{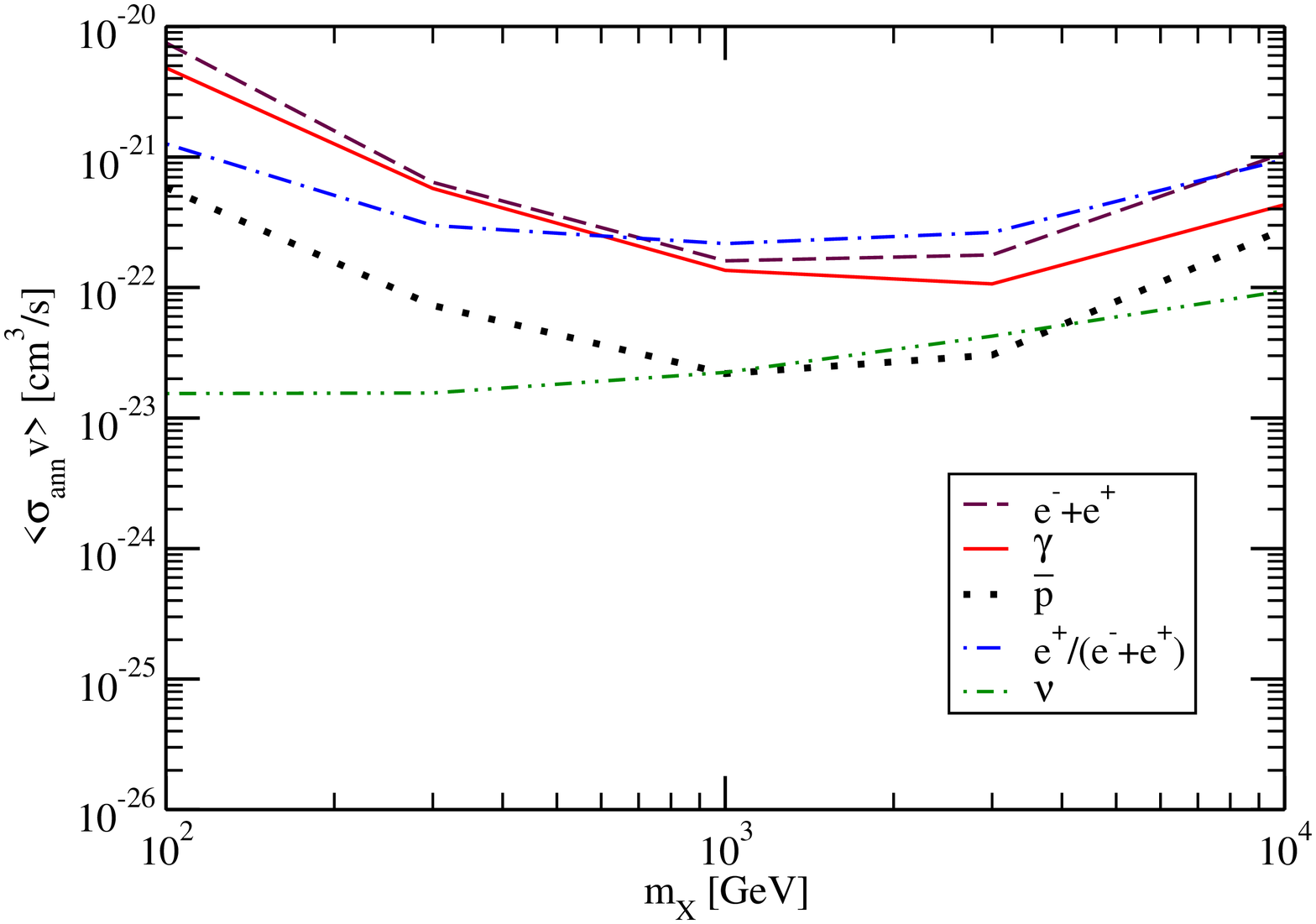}
\caption{In the top (bottom) panel, we report the exclusion plot in the  $\sigv$ vs. $m_X$ plane for the case of electron (neutrino) final states at tree level. Regions above the different curves are excluded by the labelled CR observables as described in the text.}\label{fig:constr}
\end{figure}

In Fig.~5 we summarize our bounds. Not surprisingly, for the case of an electron final state
at tree level (top panel), the bounds from electron-related observables dominate at low energies. In particular, around $m_X\sim 100\,$GeV,  the positron fraction is saturated by a cross section less than  one order of magnitude larger than the fiducial 
 value for a thermal relic, $\sigv\simeq 3\times 10^{-26}\,$cm$^3$/s. Equivalently, we expect a DM contribution above the 10\% level in some bins of the
PAMELA data, even in absence of astrophysical boost factors due to a clumpy DM distribution. 
Note also that  for the same mass range  the diffuse gamma-ray bounds fall, despite the $\alpha/\pi$ 
suppression, within two orders of magnitude of 
the fiducial value, confirming the promising role of this channel for 
detection in case
of  ``ordinary'' final state br's and of more targeted searches (see e.g.~\cite{Serpico:2009vz}
 and refs. therein). 
Above $m_X\sim $\,TeV,  the limits derived from different channels have 
roughly the same strength. Hence saturating  the  electron flux by DM
annihilations will lead to tensions with the antiproton bound,
and eventually with the gamma-ray and neutrino bounds too.

For the neutrino final state the best limit comes from neutrino observations, but for 
the region of a few TeV where the antipronton fraction provides a better  constraint.
Also, in the same range  the other channels lead to constraints looser by less than one
order of magnitude, with the gamma-ray being the most competitive one. 

In deriving the above bounds, we assumed conservatively that the products of the electron energy losses are not observed otherwise, while it is known that they can lead to potentially interesting
signatures in ``softer'' gamma-rays via inverse Compton scattering, for example. Also, we ignored
gamma-ray observations closer to the GC, while it is estimated that gamma-ray constraints at intermediate galactic latitudes or towards the GC are stronger than the ones
presented here (see e.g.~\cite{Dodelson:2007gd,Serpico:2008ga,Baltz:2008wd}.)
Finally, enforcing the HESS constraints~\cite{HESS} on the combined $e^{-}+e^{+}+\gamma$ 
flux at energies $E\approx 0.6$--4\:TeV might produce---depending on the propagation parameters---bounds 
a factor of a few more stringent than the ones shown.
Even with these caveats in mind, however, it it interesting to note that
multi-TeV scale particles annihilating
into neutrinos at tree-level might turn out to be similarly or even more  
constrained than particles annihilating into electrons.

The comparison with the electron case shows that even for the largest masses the tree-level, monochromatic neutrinos provide the leading constraint.
This is essentially due to two facts: i) The cross section for muon production and the muon range continue to increase also for $s\gg m_X^2$, partially compensating the $m_X^{-2}$ 
suppression in the differential flux.  ii) The lack of energy losses  for neutrinos, which is instead
a key factor for very energetic electrons. 

On general grounds, it appears safe to conclude that if a DM candidates in the 
few TeV mass range contributes a significant fraction of the CR 
electron/positron flux, then a non-negligible antiproton flux is produced as
well that is close to the current bound.  Especially in the light of the 
more refined measurements expected in the next years with AMS-02~\cite{AMS-02}, this is an important signature to keep in mind. Further, for the same class of candidates one expects measurable signatures in both the gamma-ray and the neutrino flux, with interesting observational perspectives from the direction of the inner Galaxy.

\section{Conclusions}\label{conclusions}

We discussed the role of electroweak bremsstrahlung for indirect 
dark matter signatures, which has been typically neglected in phenomenological 
studies heretofore. Our approach was  to calculate the branching
ratios and the fragmentation spectra $\d N_i/\d x$ of secondaries
considering electroweak radiation  only from the final state. Therefore,
our results are applicable mainly to models tailored to produce only 
leptons as final state and with DM mass in the TeV range. In particular,
our analysis applies to several models trying to match features in $e^+e^-$ CR data
with TeV-scale DM, but e.g.\ not to the benchmark case of neutralino annihilations in the MSSM
(where final states containing hadrons or gauge bosons are anyway allowed at tree-level).

An  important phenomenological consequence of electroweak bremsstrahlung
is its impact on the predicted photon and electron/positron spectra.
Secondaries from $W$ and $Z$ decays can dominate in a certain $x$ range
the spectra, leading to changes in the normalization and the shape
of the secondary spectra. 
Thus it is mandatory to assess the importance of these ``corrections'' in 
a specific model under consideration, before one attempts to fit cosmic ray
or photon data. Models viable at tree level may be ruled out by the 
more realistic spectrum or, vice versa, large boost factors required 
might be moderately loosened.

On more general grounds, at least in cases where no new light particles are introduced in the spectrum, it appears that DM candidates in the  few TeV mass range contributing a significant fraction of the CR electron/positron flux should lead also to a non-negligible antiproton flux, close to the current bound. Additionally, for the same class of candidates we expect  measurable signatures in both the gamma-ray and the neutrino flux. These ``multimessenger'' signatures are not surprising and have often been implicitly assumed in past phenomenological works on indirect DM detection. However, it is interesting that the conclusion holds virtually unchanged also for heavy DM candidates engineered to produce only leptons as tree level final states.

\acknowledgments
We are grateful to J.~Alwall and T.~Stelzer for advice on Madgraph,
and especially to David Grellscheid for help using HERWIG++.

\vskip0.5cm
{\it Note added:} 
During the preparation of this work, some of the  considerations developed 
here have been mentioned (in a different context) in the model presented in
Ref.~\cite{Falkowski:2009yz}.


\end{document}